\PassOptionsToPackage{colorlinks=true, citecolor=blue, urlcolor=blue, linkcolor=black}{hyperref}
\documentclass{article}
\usepackage{iclr2026_conference,times}

\usepackage{amsmath,amsfonts,bm}

\def\eqref#1{equation~\ref{#1}}

\def\1{\bm{1}}

\DeclareMathAlphabet{\mathsfit}{\encodingdefault}{\sfdefault}{m}{sl}
\SetMathAlphabet{\mathsfit}{bold}{\encodingdefault}{\sfdefault}{bx}{n}

\usepackage{graphicx}
\usepackage{booktabs}
\usepackage{caption}
\usepackage{subcaption}
\usepackage{amsmath}
\usepackage{amssymb}
\usepackage{url}
\usepackage{multirow}
\usepackage{siunitx}
\usepackage{threeparttable}
\usepackage[table]{xcolor}
\usepackage{pifont}
\usepackage{float}
\usepackage[skip=5pt]{caption}

\usepackage{fontawesome5}
\usepackage[T1]{fontenc}
\usepackage{twemojis}

\usepackage{hyperref}
\usepackage{cleveref}

\newcommand{\modelraw}{QuarkAudio}

\iclrfinalcopy

\begin{document}

\begin{center}
	\vspace*{-3.5em}
	
	{\LARGE \textbf{QuarkAudio Technical Report}}
	
	\vspace{0.8em}
	
	\normalsize \bfseries
	Chengwei Liu\textsuperscript{1,$\star$} \quad
	Haoyin Yan\textsuperscript{2,*} \quad
	Shaofei Xue\textsuperscript{1,2,$\dagger$} \quad
	Xiaotao Liang\textsuperscript{1} \quad
	Xiaofu Chen\textsuperscript{1,3} \\[3pt]
	Bin Gong\textsuperscript{1,3} \quad
	Zheng Xue\textsuperscript{1} \quad
	Gang Song\textsuperscript{1}
	
	\vspace{0.6em}
	
	\normalsize
	\textsuperscript{1}Intelligent Connectivity, Alibaba Group \quad
	\textsuperscript{2}Tongyi AI Lab, Alibaba Group \quad
	
	\textsuperscript{3}Zhejiang University \\[4pt]
	$^\star$Equal contribution \quad
	$^\dagger$Corresponding author
	
	\vspace{0.8em}
	
	\scriptsize
	\renewcommand{\arraystretch}{1.2}
	\setlength{\tabcolsep}{1.5pt}
	\begin{tabular}{@{}l@{\hspace{2mm}}l}
		\faGithub\ \textbf{Code:}     & \url{https://github.com/alibaba/unified-audio} \\
		\includegraphics[height=1.8ex]{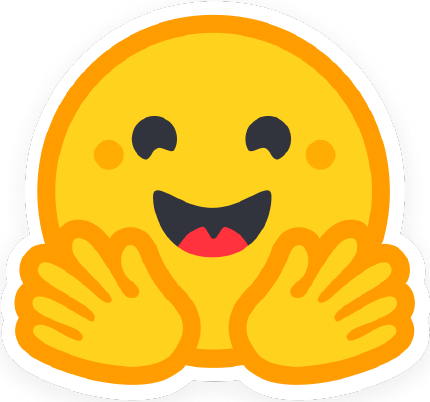}\ \textbf{Model:} & \url{https://huggingface.co/QuarkAudio} \\
		\faMicrophone\ \textbf{Demo:} & \url{https://alibaba.github.io/unified-audio} \\
	\end{tabular}
\end{center}




\begin{abstract}

Many existing audio processing or generation models involve task-specific architectures, 
resulting in fragmented development efforts and limited extensibility. 
It is promising to design a unified framework that has ability to address multiple tasks, 
simultaneously providing robust instruction and audio understanding while delivering high-quality audio generation.
This requires a highly compatible paradigm design, 
powerful modeling capabilities of backbone, and a high-fidelity audio recovery module. 
To meet the above requirements, this technical report introduces \textbf{\modelraw{}}, a novel decoder-only autoregressive (AR) LM-based generative framework to unify multiple tasks.
The framework includes a unified discrete audio tokenizer \textbf{H-Codec}, 
which integrates self-supervised learning (SSL) representation within the audio tokenization and reconstruction process. 
Some innovative improvements for \textbf{H-Codec} are proposed, such as dynamic frame-rate mechanism and extending audio sampling rate to 48 kHz.
\textbf{\modelraw{}} unifies tasks by taking task-specific conditional information as the conditioning sequence of decoder-only LM, 
and the discrete token of target audio is predicted in an AR manner. 
The framework supports a wide range of audio processing and generation tasks, including: speech restoration (SR), 
target speaker extraction (TSE), speech separation (SS), voice conversion (VC), and language-queried audio source separation (LASS).
In addition, we extend the downstream tasks to universal free-form audio editing guided by natural language instructions (including speech semantic editing and audio event editing).
Experimental results demonstrate that \textbf{H-Codec} achieves remarkable audio reconstruction quality with a low frame rate, 
improving both the efficiency and performance of downstream audio generation, and \textbf{\modelraw{}} achieves competitive or comparable performance 
in comparison with state-of-the-art task-specific or multi-task systems across multiple tasks.

\end{abstract}

\section{Introduction}
\label{sec:intro}

The development of general-purpose large models capable of unifying diverse tasks has emerged as a pivotal trend in artificial intelligence, driven by the demand for efficient and scalable solutions across modalities. 
In the realms of text and vision, foundational models such as GPT and CLIP have demonstrated remarkable success in multi-task learning frameworks, enabling cross-modal alignment and task-agnostic inference. 
However, extending this paradigm to audio, a modality characterized by complex temporal dynamics and heterogeneous task requirements, remains a significant challenge.
Current audio-specific models often exhibit task fragmentation, where specialized architectures are required for distinct objectives such as speech synthesis or sound event detection and localization. 
This fragmentation stems from three critical bottlenecks: the difficulty in extracting audio representations that generalize across tasks with varying semantic and acoustic requirements, 
the lack of a unified framework capable of harmonizing diverse input-output modalities (e.g., text-to-speech, speech-to-music, or audio editing), 
and the inherent conflicts between task-specific constraints, such as the trade-off between high-fidelity reconstruction in generation tasks and real-time efficiency in interactive applications. 
These challenges hinder the progress toward a cohesive audio foundation model that mirrors the versatility observed in text and vision domains.

Neural audio codecs utilize neural networks to obtain highly compressed discrete representations of audio waveforms and 
aim to reconstruct high-fidelity signal form discrete tokens. 
Acoustic tokens, such as SoundStream~\citep{soundstream} utilizes residual vector quantization (RVQ) where 
each quantizer refines the residuals left by the previous one, 
obtaining parallel multi-layer tokens and achieving remarkable reconstruction quality. Many works including Encodec~\citep{encodec} and DAC~\citep{dac} follow this paradigm to improve performance. 
These tokens capture fine-grained acoustic characteristics of raw waveforms, offering an alternative to conventional frame-level features such as mel-spectrograms. 
The development of acoustic codecs primarily encompasses the following directions:
Model scale-up, as exemplified by StableCodec\,\citep{Julian2024StableCodec} and WavTokenizer\,\citep{Ji2024WavTokenizer};
Variable frame lengths, represented by SNAC\,\citep{Hubert2024SNAC}. 

Conversely, semantic tokens, such as HuBERT, Whisper, and CosyVoice \citep{hubert, Whisper, du2024cosyvoice}, 
are rich in semantic content but tend to lose acoustic details.
With the development of LM, the research focus of codecs has gradually shifted 
from reducing data transmission costs toward the integration with LM, which ensures the high quality of generated audio. 
This requires codecs~\citep{semanticodec,moshi} to preserve more semantic information that can be understood and modeled by LM. 
X-Codec~\citep{xcodec} integrates the representations from the pre-trained SSL model to enhance semantic preservation, 
improving both reconstruction quality and downstream TTS performance. 
Some studies~\citep{wavtokenizer,unicodec} explore single-layer codecs that are more suitable for autoregressive modeling in LM. 
X-Codec2~\citep{xcodec2} utilizes finite scalar quantization (FSQ)~\citep{fsq} to perform single-layer quantization, 
enlarging the code space. BiCodec~\citep{SparkTTS} generates a hybrid token stream combining semantic and global tokens, 
which are derived from a SSL model and a speaker verification model, respectively. 
However, single-layer codecs with a low frame rate still faces challenges in high-fidelity reconstruction~\citep{xcodec2}, e.g., speaker similarity. 

In practice, downstream LMs are capable of generating multi-layer tokens in parallel~\citep{musicgen, t5tts}, 
thereby relaxing the requirement for single-layer quantization. 
This paradigm relies more heavily on the modeling capacity of LMs, raising the upper bound of the codec's reconstruction capability.
In this context, the frame rate of codecs plays a critical role, which determines the number of time steps for inference. 
To address the above issues, we propose \textbf{H-Codec}, a dual-stream audio codec involving acoustic and semantic branch is developed for high-fidelity waveform reconstruction.
Our H-Codec benefits from the RVQ technique and SSL representations, achieving significant reconstruction quality in the domain of speech, music, and general audio. 
The low frame rate ensures efficient generation when integrated with our \modelraw{} framework.

Creating an unified framework that can tackle diverse tasks stands as a critical research goal in the field of artificial intelligence. For the audio understanding task, KimiAudio \citep{KimiTeam2025KimiAudioTR} proposes an audio tokenizer that concatenates the semantic audio token with continuous acoustic vectors from Whisper \citep{Whisper} encoder to enhance perception capability and output a discrete semantic token. DualSpeechLM \citep{Wang2025DualSpeechLMTU} proposes a dual token modeling paradigm, using a dedicated USToken for input and an acoustic token for output. Many studies integrate generative modeling into audio tasks in recent years. 
For the SR task, SELM~\citep{wang2024selm} applies k-means to quantize noisy speech representations obtained  by WavLM~\citep{chen2022wavlm} into discrete tokens, and then a Transformer-based speech LM maps the noisy tokens to clean tokens.
For the LASS task, FlowSep~\citep{yuan2025flowsep} learns linear flow trajectories from noise to target source features within the variational autoencoder (VAE) latent space, 
which are guided by the encoded text embeddings and the mixture audio. 
For the speech edit task, VoiceBox \citep{Le2023VoiceboxTM} introduced a non-autoregressive editing paradigm based on flow matching. However, its reliance on alignment models like MFA to specify the editing region adds operational overhead in practical scenarios. 
InstructSpeech \citep{huang2024instructspeech} represents a significant advancement, employing triplet data construction, multitask learning, and multistep reasoning to enable instruction-based editing. 

Compared to discrete audio codec based method, especially decoder-only AR models which can elegantly 
integrate conditional information as a prefix sequence, continuous methods usually requires complex design to combines multimodal conditions, limiting the extensibility to more tasks. 
In addition, discrete audio representation plays an important role in combining with LLM~\citep{qwen3omni}, 
bridging the natural language instructions and continuous waveform. 
Therefore, we develop a decoder-only AR LM-based framework (\textbf{\modelraw{}}) to unify audio tasks. 
It utilizes continuous conditional embeddings to maximize the preservation of semantic and acoustic information, 
predicting multi-layer codec tokens which reduce the quantization loss.

In terms of speech representations, understanding and generation tasks impose fundamentally different demands: 
understanding benefits from compact, semantics-focused encodings, while generation requires rich, 
high-quality acoustic details\citep{Inclusion2025MingOmniAU,yan2025minguniaudio}. As a result, 
most existing speech language models resort to one of two architectural compromises: 
either maintaining separate representations for understanding and generation \citep{Xu2025Qwen3OmniTR,KimiTeam2025KimiAudioTR}, 
or relying on discrete tokens for both \citep{defossez2024moshi, coreteam2025mimoaudio,Huang2025StepAudioUU}.
The challenge of the former approach lies in unifying the two representation schemes into a single modeling space \citep{Le2023VoiceboxTM}, 
while the performance of the latter method is constrained by information loss due to discretization.
Additionally, audio generation models still face challenges in terms of audio quality and generalization ability across tasks. 
This fragmentation results in redundant development efforts, inconsistent performance, and limited extensibility.

Some studies aim to unify multiple tasks within a single framework, including AnyEnhance~\citep{zhang2025anyenhance}, 
UniAudio~\citep{yang2024uniaudio}, LLaSE-G1~\citep{kang2025llaseg1}, UniSE~\citep{unise}, and Metis~\citep{wang2025metis}. 
These methods utilizes the LM backbone combined with discrete audio codec and 
exhibit remarkable generative ability, which benefit from the semantic understanding and contextual modeling capabilities of LMs.
However, challenges still exist in terms of audio quality and generalization ability across tasks. 
For instance, few unified models are capable of handling the SS task, as it generally requires customized 
architecture to output multi-track speech. 
In terms of speech tasks, free-form audio editing—particularly guided by natural language instructions—is an emerging and highly challenging direction. Unlike conventional audio tasks that operate on fixed input-output mappings, audio editing requires deep understanding of complex linguistic semantics, fine-grained control over acoustic attributes, and high-fidelity generation~\citep{yan2025minguniaudio,Inclusion2025MingOmniAU,google2025gemini}. For instance, users may wish to ``change the speaker's emotion from angry to calm'' or ``add birdsong in the background.'' Such capabilities demand joint modeling of audio semantics, prosody, and environmental context. While pioneering works like WavCraft~\citep{liang2024wavcraft}, SAO-Instruct~\citep{instruct}, Voicecraftx~\citep{zheng2025voicecraftx}, Step-Audio-EditX~\citep{yan2025step-audio}, and Ming-UniAudio~\citep{yan2025minguniaudio} have begun exploring this space, existing solutions still face significant limitations in handling compositional instructions and preserving long-term audio coherence.
To reconcile the issue of competing requirements of representations on understanding and generation tasks, 
we utilize continuous representations for robust semantic modeling and discrete representations for high-fidelity audio synthesis, with a unified LLM harmonizing the interaction between these two representation spaces.

The contributions of this work can be summarized as follows:
\begin{enumerate}
	\item \textbf{Unified Discrete Audio Tokenizer}: We present \textbf{H-Codec}, 
	which integrates self-supervised learning (SSL) representation within the audio tokenization and reconstruction process. 
	The features from waveform and SSL model are individually quantized, resulting dual-stream (acoustic and semantic) codec tokens. 
	H-Codec achieves remarkable audio reconstruction quality with a low frame rate, 
	improving both the efficiency and performance of downstream audio generation.
	We further present two advanced versions of \textbf{H-Codec}: H-Codec-1.5 introduces a dynamic frame rate mechanism built upon H-Codec-1.0, enabling adaptive temporal resolution for varying
	content complexity; H-Codec-2.0 extends the sampling rate from 16kHz to 48kHz under a fixed frame rate, significantly improving audio fidelity and high-frequency detail preservation.

	\item \textbf{Unified Audio Language Model for Generation}: \textbf{\modelraw{}} unifies tasks by taking task-specific conditional information as the conditioning sequence of decoder-only LM, 
	and the discrete token of target audio is predicted in an AR manner. 
	We utilize a special task token to distinguish different learning patterns of multiple tasks. 
	Note that our model handles diverse tasks using a single set of shared weights, thereby eliminating the need for task-specific weight adaptation.

	\item \textbf{Multi-tasks with High-Fidelity Generation:} 
	\textbf{\modelraw{}} supports a wide range of audio processing and generation tasks within a single shared architecture, and enables comprehensive audio editing based
	on user instructions and available audio clips, covering both semantic content modifications and audio event editing. In addition, this framework achieves high-fidelity generation quality 
	in terms of SR, TSE, SS, VC, LASS, and EDIT tasks, demonstrating strong competitiveness compared to 
	state-of-the-art (SOTA) task-specific or multi-task baselines.

\end{enumerate}

\section{Model Architecture}
\begin{figure*}[t]
	\centering
	\includegraphics[width=1.0\textwidth]{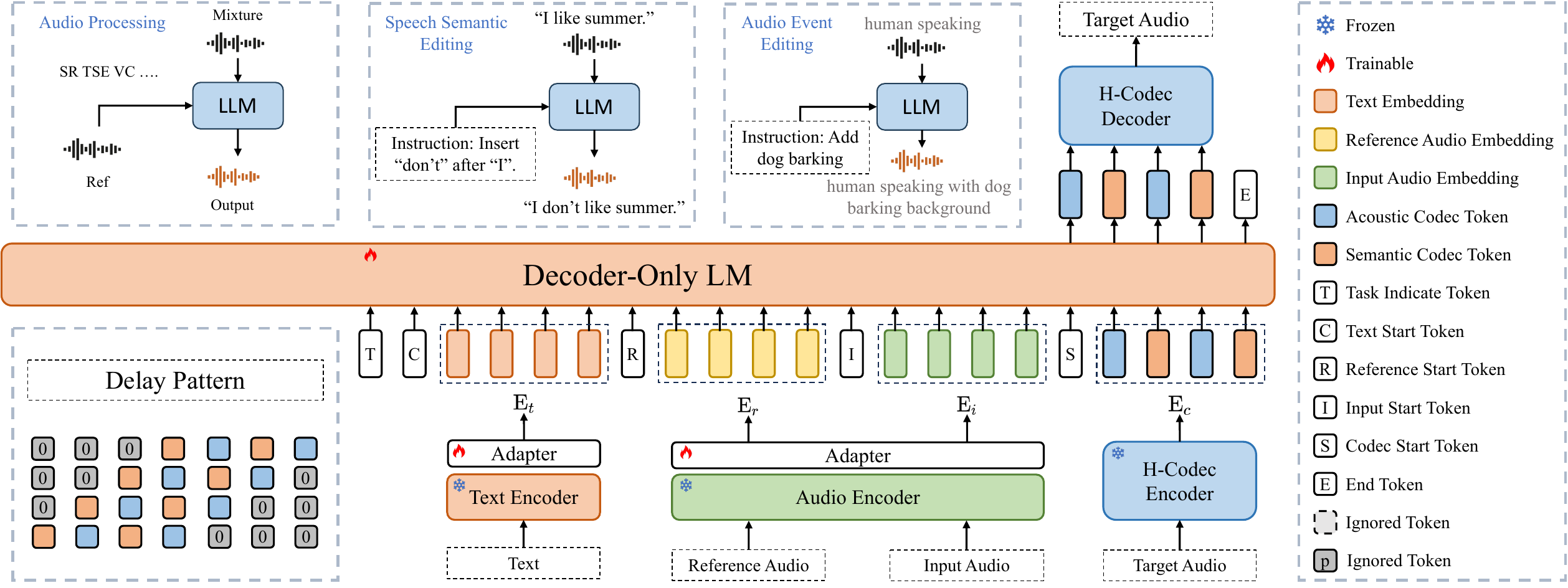}
	\caption{The overall architecture of \modelraw{}, which is a straightforward model for multiple audio tasks. For simplicity, we illustrate the AR process with single-layer codec tokens and it actually operates in a multi-layer AR manner with delay pattern.}
	\label{fig:QuarkAudio}
\end{figure*}

As shown in \Cref{fig:QuarkAudio}, \textbf{\modelraw{}} is a unified, 
autoregressive LM-based audio generation framework comprising four key components: 
(i) a novel dual-stream \textbf{H-Codec}; (ii) a text encoder with adapter; (iii) an audio encoder with adapter; (iv) a decoder-only LM backbone.

\subsection{Unified Discrete Tokenizer}
\begin{figure}[h]
	\centering
	\includegraphics[width=1.0\textwidth]{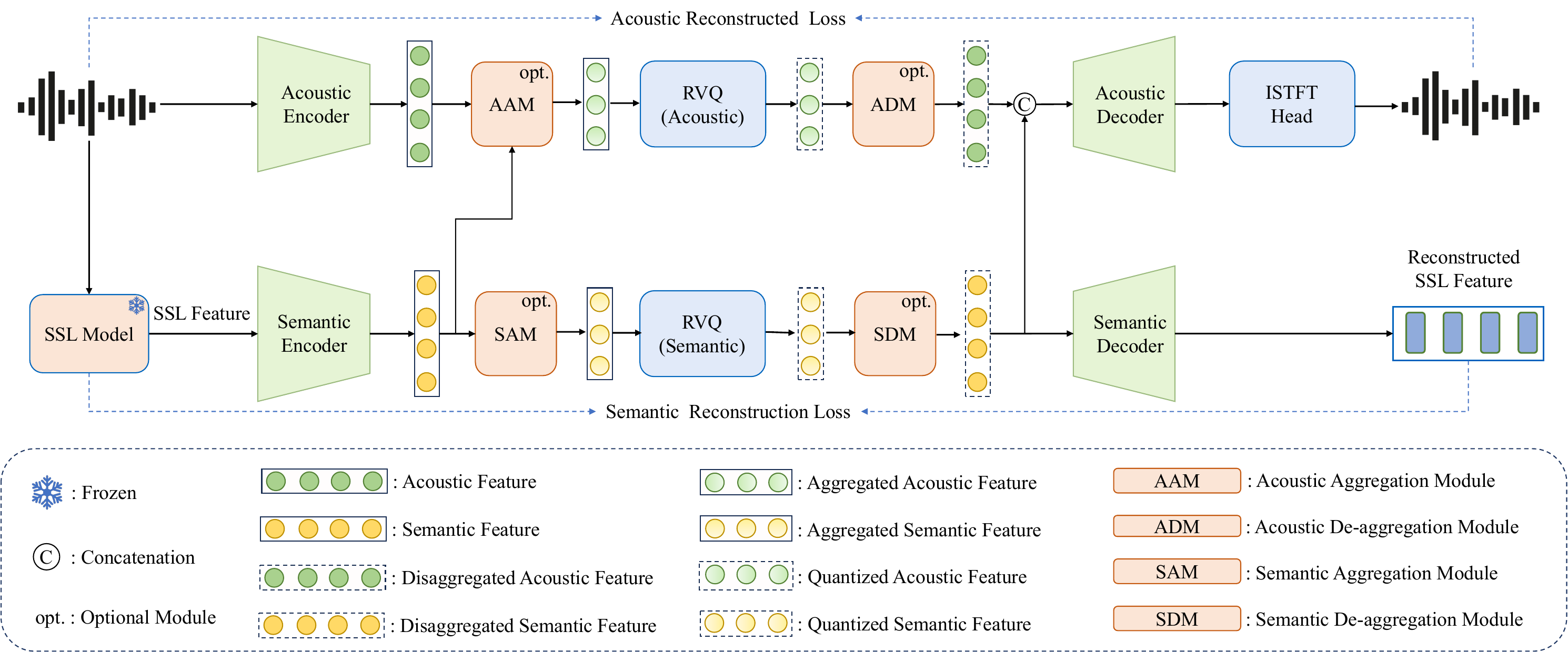}
	\caption{Architecture of H-Codec}
	\label{fig:H-Codec}
\end{figure}

Guided by the aforementioned design thoughts, we proposed H-Codec, as shown in \Cref{fig:H-Codec}. 
Following the general architecture of audio codec, H-Codec is built upon three core elements: codec encoder, quantizer module and codec decoder.
Inspired by X-Codec~\citep{xcodec}, we incorporate pretrained models to facilitate the preservation of semantic information. 
However, unlike X-Codec, which fuses acoustic and semantic information and then quantizes the combined representation using a single codebook, 
we employ separate codebooks to quantize the two types of features independently, leading to dual-stream codec tokens. 
We extend the original design of H-Codec (``H-Codec-1.0'') in UniTok-Audio~\citep{unitok_audio} to H-Codec-1.5 
and H-Codec-2.0, where the former adopts a dynamic frame-rate mechanism to reduce frame rate, 
and the latter expands audio sampling rate from 16 kHz to 48 kHz, 
supporting higher-fidelity audio reconstruction and broader application scenarios such as high-quality multimedia content.

\subsubsection{Encoder and Decoder}
In the encoding stage, the raw waveform $\boldsymbol{x} \in \mathbb{R}^{n}$ is fed into the acoustic encoder to extract frame-level acoustic features, 
where $n$ represents the number of waveform samples. 
H-Codec-1.5 inherits the module architecture of H-Codec-1.0, which is illustrated in UniTok-Audio. 
H-Codec-2.0 upgrades the architecture of acoustic encoder. 
We perform short-time Fourier transform (STFT) and 
concatenate the magnitude and phase spectrum as input feature, 
which is then fed into the stack of ConvNeXtBlock layers and Transformer layers. 
A RVQ~\citep{soundstream} quantizer is utilized to quantize acoustic features.
Synchronously, a pre-trained SSL model 
(speech-specialized WavLM\footnote{https://huggingface.co/microsoft/wavlm-base-plus}~\citep{chen2022wavlm} for H-Codec-1.5 and 
audio-generalist HuBERT\footnote{https://huggingface.co/bosonai/hubert\_base}~\citep{hubert} for H-Codec-2.0) 
extracts SSL features by averaging outputs from all transformer layers
and the quantized semantic feature is obtained by applying the semantic encoder and RVQ quantizer. 

For the decoding stage, 
the quantized acoustic and semantic features are concatenated along the hidden dimension, and the 
waveform is reconstructed by utilizing acoustic decoder and the inverse short-time Fourier transform (ISTFT) head following Vocos~\citep{vocos}.
We believe that decoupling acoustic and semantic features enables each branch to learn distinct representations,
which is beneficial for improving reconstruction quality.
Additionally, the quantized semantic feature is processed by the semantic decoder to reconstruct the SSL feature. 
This ensures that the quantized semantic features retain sufficiently rich semantic information. 




\subsubsection{Dynamic Frame Aggregation and De-aggregation}
Inspired by the dual-branch codec framework FlexiCodec~\citep{liFlexiCodecDynamicNeural2025}, 
we implement a dynamic frame-rate mechanism based on semantic feature similarity 
in H-Codec (i.e. H-Codec-1.5). 
Specifically, the system comprises aggregation and de-aggregation modules for both the acoustic and semantic branches. 

The semantic aggregation module (SAM) first computes the cosine similarity scores between all adjacent semantic features, 
where the semantic feature sequence is extracted by the semantic encoder. 
It then scans through the similarity sequence to identify multiple 
consecutive segments in which all similarity scores exceed a predefined threshold. 
For each identified segment, the frames are subsequently mapped into a single feature representation via a local attention mechanism.
The acoustic aggregation module (AAM) inherits the segment boundaries determined by the semantic aggregation module to guide the compression of acoustic features. 
This design enables semantic-guided acoustic feature compression while ensuring temporal consistency across the feature dimensions. 
Furthermore, to facilitate the restoration of the original frame rate during decoding, 
we follow the design of \textit{VARStok}~\citep{zhengSayMoreLess2025} by embedding the segment length into the final code index as follows:
\begin{equation}
	\mathbf{c}_{i}^{\prime} = (d_i - 1) \times K + \mathbf{c}_{i},
\end{equation}
where $K$ denotes the size of the original codebook, 
$d_i$ the duration of segment~$i$, and $\mathbf{c}_{i}$ 
the original code index, respectively.


The semantic de-aggregation module (SDM) and acoustic de-aggregation module (ADM)
take as input the compressed and quantized features, 
restoring the frame rate according to the existing segment partitions. 
Specifically, the segment length is decoded from the code index by 
\begin{equation}\label{eq:decode_len}
	d_i = \left\lfloor \frac{\mathbf{c}_{i}^{\prime}}{K} \right\rfloor + 1,
\end{equation}
where $\lfloor \cdot \rfloor$ denotes the floor operation. 
The frame sequences are reconstructed by repeating the segment length to recover the original frame rate. 
Finally, to more naturally regenerate the original features, 
the repeated segments undergo a local attention module before being delivered to the decoder.
The implementation of local attention is consistent with that in FlexiCodec.

\subsubsection{Optimization Strategy}

\begin{figure}[h]
	\centering
	\includegraphics[width=0.8\textwidth]{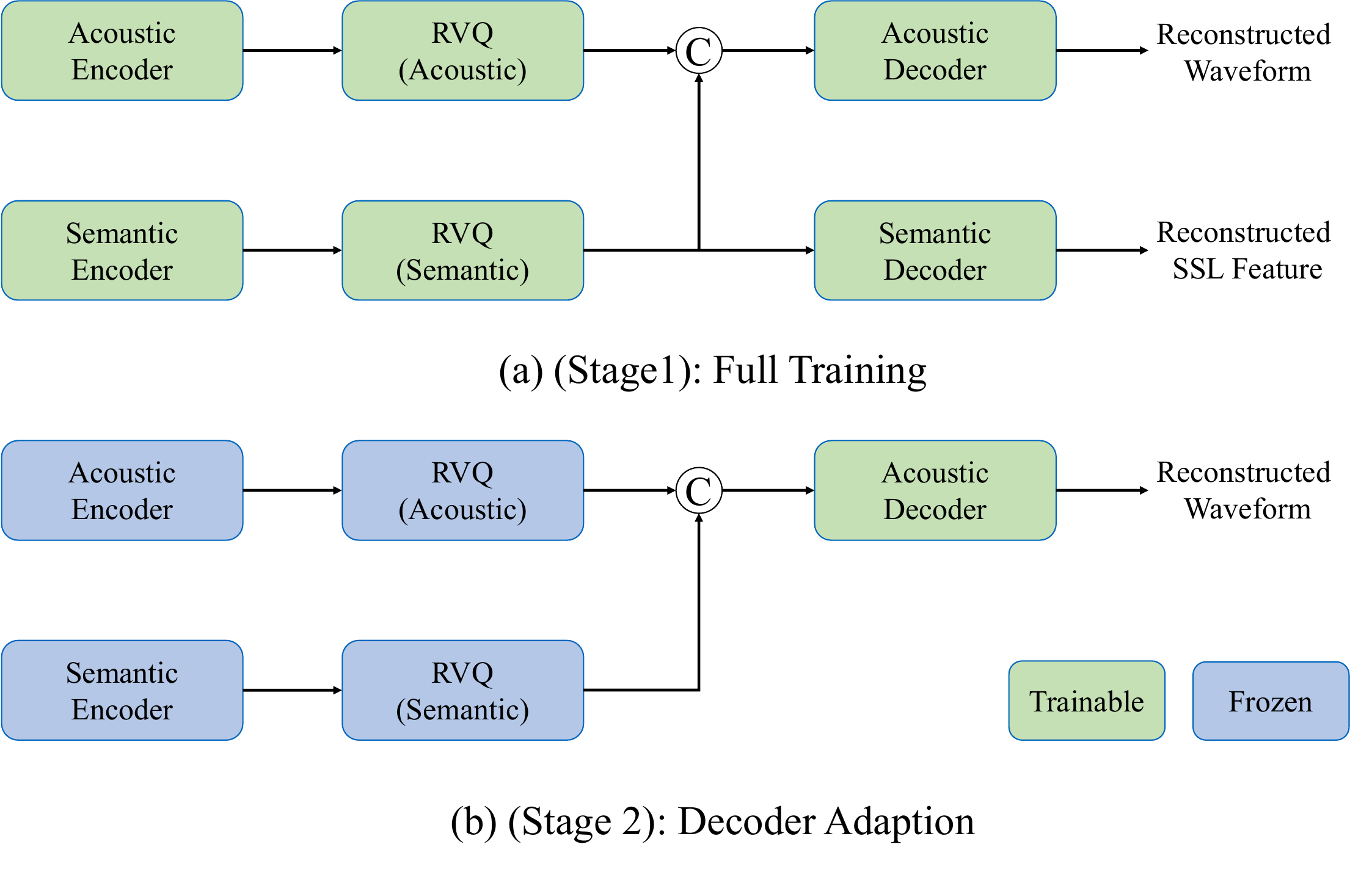}
	\caption{H-Codec-2.0 training pipeline}
	\label{fig:tokenizer_pipeline}
\end{figure}

The configurations of discriminators and optimization objective functions 
follow that described in UniTok-Audio~\citep{unitok_audio}. 
We utilize a two-stage training strategy in H-Codec-2.0, as shown in~\Cref{fig:tokenizer_pipeline}.
In stage 1, we train all modules on a large-scale and diverse audio corpus, 
enhancing robustness of token extraction in the general audio domain. 
In stage 2, we freeze the parameters 
of encoders and quantizers
to fine-tune the acoustic decoder. 
This stage further specializes the acoustic decoder to 
reconstruct audio without adapting to the varying quantized features.

\subsection{Unified Audio Language Model}

\subsubsection{Overall Framework}
To unify various audio processing or generation tasks within a single framework, 
we extract task-specific conditional information as a conditioning sequence for the decoder-only AR backbone.
Since continuous features, typically extracted from SSL models, contain richer audio details compared to discrete representations 
and are more adaptable to varying input conditions, we extract continuous features to assemble the task-conditioning sequence.
Specifically, we utilize T5-base\footnote{https://huggingface.co/google/t5-v1\_1-base}~\citep{t5} as the 
text encoder to extract embedding from audio caption. 
The same HuBERT used in H-Codec is adopted to extract continuous features from audio waveforms.
Two linear layers serve as two adapters to map the text embedding and audio features into a representation space amenable to LM AR modeling, respectively.
Given text and audio embeddings as conditions, 
we utilize LLaMA architecture~\citep{LLaMA} to predict discrete tokens of target waveform in an AR manner. 
Since the H-Codec tokens involve multiple layers, 
we apply the delay pattern~\citep{musicgen} to arrange our tokens for the trade-off between performance and computational cost. 
The detail of the paradigm can be found in~\citep{unitok_audio}.
Finally, the H-Codec decoder reconstructs high-fidelity audio from the predicted token sequence.

\begin{table}[tt]
	\centering
	\caption{Operational modes and corresponding conditions in \modelraw{}.}
	\label{tab:conditioning}
	\small
	\begin{tabular}{@{}lcc@{}} 
		\toprule
		\textbf{Mode} & \textbf{Task Token} & \textbf{Conditions} \\
		\midrule
		\midrule
		SR & ${\rm T_{SR}}$ & Degraded Speech \\
		TSE & ${\rm T_{TSE}}$ & Reference Speech, Mixture Speech \\
		rTSE & ${\rm T_{rTSE}}$ & Reference Speech, Mixture Speech \\
		VC & ${\rm T_{VC}}$ & Reference Speech, Source Speech \\
		LASS & ${\rm T_{LASS}}$ & Caption, Mixture Audio \\
		EDIT-S & ${\rm T_{EDIT_S}}$ & Instruction, Source Speech \\
		EDIT-A & ${\rm T_{EDIT_A}}$ & Instruction, Source Audio \\
		\bottomrule
	\end{tabular}
\end{table}

\subsubsection{Unifying Tasks with Operational Modes}
Following our previous work~\citep{unise,unitok_audio}, we introduce special task tokens to distinguish between different operational modes. 
To unify seven tasks (i.e., SR, TSE, SS, VC, LASS, EDIT-S, and EDIT-A), we utilize seven modes, as shown in Table~\ref{tab:conditioning}. 
The first five tasks are defined in UniTok-Audio, 
while the EDIT-S and EDIT-A denote speech semantic editing and audio event editing, respectively. 
EDIT-S performs insertion, deletion, or substitution operations on the semantics of input speech guided by textual instructions. 
EDIT-A conducts caption-based insertion or deletion of audio event elements (e.g., drumbeats, bird calls) in the input audio without the assistance of reference audio. 
Each mode corresponds to a special token and different task-specific condition types, which serve as a conditioning sequence 
for the LM backbone to estimate the conditional probability density distribution of target discrete tokens.

\section{Experiments}

\subsection{H-Codec}
\subsubsection{Experimental Setup}
\textbf{Datasets:} 
We utilize multi-domain data to train our codec, including speech, music, and audio.
The speech samples are sourced from the VoxBox dataset~\citep{SparkTTS} and internal dataset, 
where the former comprises approximately 100k hours of speech, and 
the latter includes 2.5M hours of recordings. 
For the music domain, we utilize the FMA-full~\citep{Defferrard2017FMA}, MUSDB18-HQ~\citep{MUSDB18HQ}, 
and internal datasets, involving about 150k hours of data. 
For the audio domain, we adopt AudioSet~\citep{gemmeke2017audio}, WavCaps~\citep{wavcaps}, and 
internal datasets, including about 100k hours of recordings. 
We evaluate the speech reconstruction quality on 
LibriSpeech~\citep{Librispeech} test-clean (16 kHz) and Seed-TTS-Eval~\citep{anastassiou2024seed}\footnote{https://github.com/BytedanceSpeech/seed-tts-eval} (24 kHz). 
The reconstruction quality of music and general audio is 
evaluated on MUSDB18-HQ test set (44.1 kHz) and AudioSet eval (48 kHz) respectively.
All samples are resampled to appropriate sampling rate according to model configuration or evaluation metrics.

\textbf{Implementation Details:} 
For H-Codec-1.0 and H-Codec-1.5, 
4-layer RVQ quantizers are utilized in each stream with a codebook size of 1024 and a codebook dimension of 512. 
For H-Codec-1.5, the threshold of similarity score 
and the maximum length for aggregation are set to 0.6 and 8, and 
the local attention window spans 8 preceding and 8 succeeding tokens with 8 attention heads and 32 layers. 
For H-Codec-2.0, we perform STFT with a frame length of 1920 and a hop length of 960, resulting in 50 frames per second.
Before utilizing quantization, we reduce the frame rate to 6.25 Hz 8x downsampling. 
Each of the acoustic and semantic branches consists of 16 RVQ layers. 
During training, we randomly crop 5-second segments from audio samples. 
The network is optimized using the AdamW optimizer with an initial learning rate of 
$2 \times 10^{-4}$, which is decayed based on a cosine scheduler during 600k training steps. 

\textbf{Evaluation Metrics:} 
We utilize several metrics to measure the reconstruction quality of speech, 
including the perceptual evaluation of speech quality (PESQ), short-time objective intelligibility (STOI), 
speaker similarity (SPK-SIM) and UTMOS.
The loss on Mel-scale spectrum and STFT spectrum bettween the target audio and reconstructed audio 
are computed for general evaluation in the domain of speech, music, and audio.
Details about evaluation metrics of codec can be found in Appendix~\ref{appendix:codec_metrics}.

\textbf{Baselines:} 
We compare our codec against some state-of-the-art (SOTA) baselines, 
including DAC~\citep{dac}, Encodec~\citep{defossez2022high}, X-Codec~\citep{ye2024codec}, 
X-Codec2~\citep{xcodec2}, BiCodec~\citep{SparkTTS}, WavTokenizer~\citep{wavtokenizer}, 
UniCodec~\citep{unicodec}, FlexiCodec~\citep{liFlexiCodecDynamicNeural2025}, 
MiMO-Audio-Tokenizer~\citep{coreteam2025mimoaudio}, 
XY-Tokenizer~\citep{xytokenizer}, BigCodec~\citep{xin2024bigcodec}, 
Baichuan-Audio-Tokenizer~\citep{baichuan}, Mimi~\citep{moshi}, and GLM4-Voice-Tokenizer~\citep{glm4-voice}.


\subsubsection{Experimental Results}

\begin{table}[ht]
	\centering
	\caption{Comparison between different codec models on LibriSpeech test-clean set, 
		where \textbf{FPS} and \textbf{BPS} denote frame per second and bitrate per second, respectively. 
		\textbf{FS}, \textbf{Paras}, and \textbf{Nq} indicate the sampling rate supported by the model, 
		its parameter amount, and the number of codebook layers, respectively. 
		\textbf{Unified} indicates whether the model supports general audio or only speech. For FlexiCodec and H-Codec-1.5, 
		we calculate the average FPS and BPS on test dataset.
		}
	\label{tab:codec1}
	\resizebox{\columnwidth}{!}{
    	\begin{tabular}{l c c c c c c | *{5}{c}}
			\toprule
			\textbf{Model} & \textbf{FS} & \textbf{Unified} & \textbf{FPS} & \textbf{Nq} & 
			\textbf{BPS} & \textbf{Paras} & \textbf{PESQ}($\uparrow$) & \textbf{STOI}($\uparrow$) & \textbf{UTMOS}($\uparrow$) & \textbf{SPK-SIM}($\uparrow$) & \textbf{WER}($\downarrow$) \\
			\midrule
			Ground Truth & - & - & - & - & - & - & 4.64 & 1.00 & 4.09 & 1.00 & 2.43 \\
			\midrule
			Encodec & 24k & \ding{55} & 75 & 8 & 6000 & 15M & {2.77} & \textbf{0.94} & 3.09 & {0.89} & \textbf{2.64} \\
			X-Codec & 16k & \ding{55} & 50 & 4 & 2000 & 178M & {2.77} & 0.87 & \textbf{4.21} & 0.72 & 3.13 \\
			WavTokenizer & 24k & \ding{55} & 75 & 1 & 900 & 77M & 2.39 & 0.91 & 4.00 & 0.68 & 5.43 \\
			X-Codec2 & 16k & \ding{55} & 50 & 1 & 800 & 242M & 2.43 & 0.92 & 4.13 & 0.82 & 3.53 \\
			BiCodec & 16k & \ding{55} & 50 & 1 & 650 & 156M & 2.51 & 0.92 & 4.18 & 0.80 & 3.23 \\
			FlexiCodec & 16k & \ding{55} & 8.15 & 8 & 831 & 216M & 2.46 & 0.92 & 4.12 & 0.81 & 2.79 \\
			\midrule
			DAC & 16k & \ding{51} & 50 & 4 & 2000 & 76M & 1.42 & 0.84 & 1.83 & 0.60 & 4.32 \\
			X-Codec & 16k & \ding{51} & 50 & 4 & 2000 & 178M & 2.64 & 0.92 & 3.88 & 0.77 & 3.33 \\
			UniCodec & 24k & \ding{51} & 75 & 1 & 900 & 274M & 2.56 & 0.92 & 4.00 & 0.76 & 4.23 \\
			WavTokenizer & 24k & \ding{51} & 40 & 1 & 480 & 77M & 1.88 & 0.87 & 3.78 & 0.57 & 10.03 \\
			\midrule
			H-Codec-1.0 & 16k & \ding{51} & 25 & 4+4 & 2000 & 146M & {3.05} & \textbf{0.94} & {4.08} & {0.89} & {2.85} \\
			H-Codec-1.5           & 16k & \ding{55} & 19.54 & 4+4 & 1622 & 678M & 2.60 & 0.92 & 3.99 & 0.83 & 3.02 \\
			H-Codec-2.0 (Small)   & 48k & \ding{51} & 6.25 & 16+16 & 2000 & 236M & {2.67} & {0.93} & {3.75} & {0.89} & {3.05} \\
			H-Codec-2.0 (Large)   & 48k & \ding{51} & 6.25 & 16+16 & 2000 & 1.2B & \textbf{3.10} & \textbf{0.94} & {3.89} & \textbf{0.92} & {2.70} \\
			\bottomrule
		\end{tabular}
	}
\end{table}

\begin{table}[ht]
	\centering
	\caption{Comparison of state-of-the-art audio codecs on both SEED-ZH (Chinese) and SEED-EN (English) benchmarks. 
		\textbf{PESQ}, \textbf{SPK-SIM}, and \textbf{STOI} are reported for each language, and
		higher values indicate better performance. For FlexiCodec and H-Codec-1.5, 
		we calculate the average FPS and BPS on SEED-ZH and SEED-EN, respectively.
		}
	\label{tab:codec_seed}
	\resizebox{\columnwidth}{!}{
		\begin{tabular}{l c c c | *{3}{c} | *{3}{c}}
			\toprule
			\multirow{2}{*}{\textbf{Model}}
			 & \multirow{2}{*}{\textbf{FS}} & \multirow{2}{*}{\textbf{FPS}} & \multirow{2}{*}{\textbf{Paras}} &
			\multicolumn{3}{c|}{\textbf{SEED-ZH $\uparrow$}} &
			\multicolumn{3}{c}{\textbf{SEED-EN $\uparrow$}} \\
			\cmidrule(lr){5-7} \cmidrule(lr){8-10}
			& & & & \textbf{PESQ} & \textbf{SPK-SIM} & \textbf{STOI} & \textbf{PESQ} & \textbf{SPK-SIM} & \textbf{STOI} \\
			\midrule
			GLM4-Voice-Tokenizer         & 16k & 12.5 & - & 1.06 & 0.33 & 0.61 & 1.05 & 0.12 & 0.60 \\
			Baichuan-Audio-Tokenizer     & 16k & 12.5 & - & 1.84 & 0.78 & 0.86 & 1.62 & 0.69 & 0.85 \\
			FlexiCodec       & 16k & 7.16/8.49 & 216M & 1.88 & 0.67 & 0.86 & 2.11 & 0.75 & 0.90 \\
			XY-Tokenizer                 & 16k & 12.5 & - & 2.27 & 0.77 & 0.90 & 2.14 & 0.82 & 0.90 \\
			Mimi                         & 16k & 12.5 & 79M & 2.05 & 0.73 & 0.89 & 2.01 & 0.77 & 0.89 \\
			BigCodec                     & 16k & 80 & 159M & 2.26 & 0.81 & 0.92 & 2.22 & 0.80 & 0.91 \\
			X-Codec2                    & 16k & 50 & 242M & 2.19 & 0.80 & 0.92 & 2.37 & 0.82 & 0.93 \\
			MiMo-Audio-Tokenizer         & 24k & 25 & 1.2B & 2.71 & 0.89 & 0.93 & 2.43 & 0.85 & 0.92 \\
			\midrule
			H-Codec-1.0				& 16k		& 25 			& 146M & 2.35 & 0.80 & 0.90 & 2.22 & 0.73 & 0.90 \\
			H-Codec-1.5				& 16k     	& 18.80/20.18 			& 678M & 2.40 & 0.85 & 0.90 & 2.30 & 0.82 & 0.90 \\
			H-Codec-2.0 (Small) 	& 48k 	& 6.25 			& 236M & 2.43 & 0.86 & 0.90 & 2.25 & 0.83 & 0.90 \\
			H-Codec-2.0 (Large) 	& 48k 	& 6.25 			& 1.2B & \textbf{2.88} & \textbf{0.91} & \textbf{0.93} & \textbf{2.77} & \textbf{0.88} & \textbf{0.93} \\
			\bottomrule
		\end{tabular}
	}
\end{table}

\textbf{Speech Reconstruction Performance:} 
As reported in \Cref{tab:codec1} and \Cref{tab:codec_seed}, 
our proposed H-Codec exhibits competitive signal reconstruction quality 
(PESQ and STOI), speech naturalness (UTMOS), speaker consistency (SPK-SIM), and 
semantic information preservation (WER). 
H-Codec-1.5 reduces the frame rate of H-Codec-1.0 by introducing the dynamic frame-rate mechanism 
at the expense of slight performance decay, 
which enables adaptive temporal resolution for varying content complexity.
Since multi-layer tokens can be predicted simultaneously within a single time step in downstream audio LM, 
we argue that frame rate is more critical, as the number of time steps significantly affects computational cost.
H-Codec-2.0 further reduces FPS to 6.25 Hz and extends the sampling rate to 48 kHz, supporting more application scenarios. 
We provide the small version and large version of H-Codec-2.0 for comparison, and 
the latter exhibits impressive performance in terms of most metrics, 
indicating the effectiveness of increasing model size.

\begin{table}[ht]
	\centering
	\caption{Comparison between different codec models on speech (LibriSpeech test-clean), 
		music (MUSDB18-HQ test), and audio (AudioSet eval) domain in terms of Mel loss and STFT loss.}
	\label{tab:codec2}
	\resizebox{\columnwidth}{!}{
		\begin{tabular}{l c *{6}{c}}
			\toprule
			\multirow{2}{*}{\textbf{Model}} & \multirow{2}{*}{\textbf{BPS}} & \multicolumn{2}{c}{\textbf{Speech}} & \multicolumn{2}{c}{\textbf{Music}} & \multicolumn{2}{c}{\textbf{Audio}} \\
			\cmidrule(lr){3-4} \cmidrule(lr){5-6} \cmidrule(lr){7-8}
			& & \textbf{Mel loss}($\downarrow$) & \textbf{STFT loss} ($\downarrow$) & \textbf{Mel loss}($\downarrow$) & \textbf{STFT loss} ($\downarrow$) & \textbf{Mel loss}($\downarrow$) & \textbf{STFT loss} ($\downarrow$) \\
			\midrule
			\midrule
			DAC  & 2000 & 0.6436 & 0.1667 & 0.8443	& 0.2308 & 1.9054 & 0.5164 \\
			X-Codec  &2000 & 0.4225	& 0.1161	& 0.6403	& 0.1804	& 1.5073	& 0.4193 \\
			UniCodec & 900 & 0.4147	& 0.1201 & 0.6488	& 0.1999	& 1.5403	& 0.4760 \\
			WavTokenizer & 480 & 0.5143	& 0.1364	& 0.8174	& 0.2270	& 1.8912	& 0.5201 \\
			\midrule
			H-Codec-1.0  & 2000 & 0.3315 & \textbf{0.1017} & 0.5157  & 0.1667 & 1.2312 & 0.4023 \\
			H-Codec-2.0 (Small) & 2000 & 0.3548 & 0.1069 & 0.5298 & 0.1782 & 1.3542 & 0.4218 \\
			H-Codec-2.0 (Large)   & 2000 & \textbf{0.3221} & 0.1056 & \textbf{0.5035}  & \textbf{0.1526} & \textbf{1.1125} & \textbf{0.3915} \\
			\bottomrule
		\end{tabular}
	}
\end{table}

\textbf{Audio Reconstruction Performance:} 
\Cref{tab:codec2} presents a comprehensive comparison of audio codec models on speech, music, and general audio domains. 
All baselines supports general audio reconstruction.
Notably, H-Codec-2.0 (Large) achieves lowest Mel loss and STFT loss on most domains, illustrating the powerful multi-domain reconstruction ability. 
This ensures the potential of H-Codec for extensive downstream tasks, including speech, music, and audio generation.

\textbf{Effectiveness of Two-stage Training:}
As reported in~\Cref{tab:two_stage}, 
we have observed that two-stage training yields better performance than single-stage training on both SEED-ZH and SEED-EN benchmarks. 
This verifys that fine-tuning the decoder helps futher improve the reconstruction quality, 
since the decoder can focus on generating audio from fixed representations


\begin{table}[h!]
	\centering
	\caption{Comparative results of different stage training on both SEED-ZH and SEED-EN benchmarks for H-Codec-2.0.}
	\label{tab:two_stage}
	\resizebox{\columnwidth}{!}{
	\begin{tabular}{ccccccccc}
		\toprule
		\multirow{2}{*}{\textbf{Strategy}} & \multicolumn{4}{c}{\textbf{ZH}} & \multicolumn{4}{c}{\textbf{EN}} \\
		\cmidrule(lr){2-5} \cmidrule(lr){6-9}
		& \textbf{PESQ↑}
		& \textbf{STOI↑}
		& \textbf{STFT loss$\downarrow$} & \textbf{Mel loss$\downarrow$} & \textbf{PESQ $\uparrow$} &\textbf{STOI$\uparrow$} & \textbf{STFT loss$\downarrow$} & \textbf{Mel loss$\downarrow$} \\
		\midrule
		stage1 & 2.69 & 0.92 & 0.151 & 0.464 & 2.61 & 0.92 & 0.158 & 0.382 \\
		stage2 & \textbf{2.88} & \textbf{0.93} & \textbf{0.125} & \textbf{0.422} & \textbf{2.77} & \textbf{0.93} & \textbf{0.126} & \textbf{0.345} \\
		\bottomrule
	\end{tabular}
	}
\end{table}

\subsection{Unified Audio Generation}

\textbf{Training Datasets:} 
For the training of audio processing tasks (including SR, 
TSE, SS, VC, and LASS), 
we adopt clean speech samples from the VoxBox~\citep{SparkTTS} dataset, including approximately 3.8k hours of data from 
LibriSpeech~\citep{Librispeech}, MLS\_English~\citep{MLS} and Emilia\_ZH~\citep{Emilia} subset. 
The noise corpus comprises approximately 460 hours of data from the DNS Challenge~\citep{DNS}, 
FSD50K~\citep{FSD50K}, WHAM!~\citep{wichern2019wham}, DESED~\citep{DESED}, DEMAND~\citep{DEMAND}, MUSAN~\citep{MUSAN}, 
DISCO~\citep{DISCO}, MUSDB18-HQ~\citep{MUSDB18HQ}, and TUT Urban Acoustic Scenes~\citep{UAS}. 
We include 60k room impulse response (RIR) samples from SLR28~\citep{slr28} to simulate reverberation. 
For LASS task, 
the captioned audio samples are sourced from WavCaps~\citep{wavcaps}, CLAP\_FreeSound~\citep{clap}, VGGSound~\citep{chen2020vggsound}, and internal dataset, 
resulting in approximately 40k hours. 
For the EDIT-S task, we utilize 
2M Chinese samples and 1M English samples from the internal dataset.
For the EDIT-A task, samples from 
AudioSet~\citep{audioset} and CLAP\_FreeSound~\citep{clap} are 
selected as background samples and audio event samples, respectively.
The data preparation pipelines are described in Appendix~\ref{appendix:data_preparation}.

\textbf{Implementation Details:} 
There are 16 layers with 16 attention heads and a hidden dimension of 1024 in the LLaMA-based LM backbone, 
resulting in 481M trainable parameters. 
Our model is trained using AdamW optimizer with 30 epochs, 
where the learning rate reaches a peak of 0.001 after 4000 warm-up steps and reduces at a decay factor of 0.98 in each epoch. 
The lengths of reference audio and input signal are set to 5 seconds for both training and inference phases. 
For the SR, TSE, SS, VC, and LASS task, we train the multi-task version for evaluation. 
For the editing task, we train task-specific versions for EDIT-S and EDIT-A, respectively. 
These downstream tasks are implemented using H-Codec-1.0.


\textbf{Evaluation Metrics:} 
We adopt multiple evaluation metrics to assess different aspects of the generated audio across tasks. 
For speech tasks, we evaluate quality by DNSMOS (SIG, BAK, OVRL) and NISQA, speaker similarity by SIM, 
intelligibility by WER, and continuity by PLCMOS. 
For audio tasks, we utilize FAD, CLAPScore, and CLAPScore$_A$ to measure the quality of generated audio.
Details about evaluation metrics can be found in Appendix~\ref{appendix:task_metrics}.

\subsubsection{Audio Processing Performance}

\textbf{Evaluation Datasets:} 
For the evaluation datasets, we evaluate speech restoration performance on 2020 DNS Challenge~\citep{DNS} test set (``With Reverb'') and 
2022 PLC Challenge~\citep{plc_challenge} blind test set for speech enhancement (SE) subtask and packet loss cancellation (PLC) subtask, respectively. 
The performance of TSE is evaluated on the Libri2Mix~\citep{librimix} clean test set. 
The SS, VC and LASS performance is evaluated on Libri2Mix noisy test set, 
VCTK~\citep{yamagishi2019cstr} dataset and 2024 DCASE LASS validation set, respectively. 

\textbf{Results:} 
The comparison between \modelraw{} and prior works is demonstrated in \Cref{tab:main_results}. 
For each task, we select one existing model whose training data is closely aligned with our setting. 
It can be seen that our model achieves comparable or superior performance to baselines.
More detailed results can be found in UniTok-Audio~\citep{unitok_audio}.


\begin{table}[tp]
	\setlength{\tabcolsep}{6pt}
	\centering
	\small
	\caption{Performance evaluation of \modelraw{} and baselines across all tasks.}
	\resizebox{\linewidth}{!}{
		\begin{tabular}{l c c c}
			\toprule
			\textbf{Task}
			& \textbf{Model}
			& \textbf{Metrics} 
			& \textbf{Results} \\
			\midrule
			
			\multirow{2}{*}{SE} 
			& LLaSE-G1~\citep{kang2025llaseg1} 
			& $SIG{\uparrow} \mid BAK{\uparrow} \mid OVRL{\uparrow}$ 
			& $3.59 \mid 4.10 \mid 3.33 $ \\
			& \modelraw{} 
			& 
			& $\mathbf{3.67} \mid \mathbf{4.12} \mid \mathbf{3.41}$ \\
			\midrule
			
			\multirow{2}{*}{PLC} 
			& LLaSE-G1~\citep{kang2025llaseg1} 
			& $OVRL{\uparrow} \mid PLCMOS{\uparrow}$ 
			& $ 3.03 \mid 3.68$ \\
			& \modelraw{} 
			& 
			& $\mathbf{3.35} \mid \mathbf{4.58}$ \\
			\midrule
			
			\multirow{2}{*}{TSE} 
			& LauraTSE~\citep{LauraTSE} 
			& $SIG{\uparrow} \mid BAK{\uparrow} \mid OVRL{\uparrow} \mid NISQA{\uparrow} \mid SIM{\uparrow}$ 
			& $3.61 \mid \mathbf{4.08} \mid \mathbf{3.34} \mid \mathbf{4.33} \mid \mathbf{0.97}$ \\
			& \modelraw{} 
			& 
			& $\mathbf{3.62} \mid 4.05 \mid 3.32 \mid 4.00 \mid 0.95$ \\
			\midrule
			
			\multirow{2}{*}{SS} 
			& LLaSE-G1~\citep{kang2025llaseg1} 
			& $SIG{\uparrow} \mid BAK{\uparrow} \mid OVRL{\uparrow}$ 
			& $3.48 \mid 3.83 \mid 3.11 $ \\
			& \modelraw{} 
			& 
			& $\mathbf{3.56} \mid \mathbf{4.04} \mid \mathbf{3.25}$ \\
			\midrule

			\multirow{2}{*}{VC} 
			& Metis-VC~\citep{wang2025metis} 
			& $WER{\downarrow} \mid SIM{\uparrow} \mid DNSMOS{\uparrow} \mid NISQA{\uparrow}$ 
			& $4.49 \mid 0.50 \mid 3.48 \mid 4.46$ \\
			& \modelraw{} 
			& 
			& $\mathbf{3.02} \mid \mathbf{0.50} \mid \mathbf{3.51} \mid \mathbf{4.51}$ \\
			\midrule
			
			\multirow{2}{*}{LASS} 
			& FlowSep~\citep{yuan2025flowsep} 
			& $FAD{\downarrow} \mid CLAPScore{\uparrow} \mid CLAPScore_A{\uparrow}$ 
			& $ \mathbf{0.50} \mid 20.00 \mid 63.47 $ \\
			& \modelraw{} 
			& 
			& $ 0.68 \mid \mathbf{28.85} \mid \mathbf{65.56}$ \\
			\midrule

			\bottomrule
		\end{tabular}
	}
	\label{tab:main_results}
\end{table}

\subsubsection{Speech Semantic Editing}

\textbf{Evaluation Datasets:} 
To evaluate speech editing capabilities, 
We randomly selected 100 utterances in SEED-ZH and 100 utterances in SEED-EN. 
For each language, we generate 300 samples for insertion, deletion, and substitution task (100 for each task) based on the corresponding 100 utterances. 
The data preparation pipeline is detailed in Appendix~\ref{appendix:data_preparation}. 

\textbf{Results:} 
Table \ref{tab:main_results_edit} summarizes the objective results on the EDIT-S task, 
and \modelraw achieves speech quality comparable to VoiceCraft in terms of OVRL, UTMOS and NISQA, 
demonstrating strong signal reconstruction fidelity and naturalness. 
However, it exhibits higher WER, which can be attributed to weaker semantic alignment in the current framework due 
to its end-to-end codec-based architecture that lacks explicit token-level modeling. 
Ongoing work focuses on refining the semantic bottleneck through auxiliary 
supervision to improve textual accuracy without compromising audio quality. 

\begin{table}[h!]
	\centering
		\caption{Performance comparison on edit task. The best results are in \textbf{bold}.}
		\label{tab:main_results_edit}
		\resizebox{\columnwidth}{!}{
			\begin{tabular}{ccccccccc}
				\toprule
				\multirow{2}{*}{\textbf{Method}} & \multicolumn{4}{c}{\textbf{Chinese}} & \multicolumn{4}{c}{\textbf{English}} \\
				\cmidrule(lr){2-5} \cmidrule(lr){6-9}
				& \textbf{WER$\downarrow$}
				& \textbf{OVRL$\uparrow$}
				& \textbf{UTMOS$\uparrow$} & \textbf{NISQA$\uparrow$} & \textbf{WER$\downarrow$} &\textbf{OVRL$\uparrow$} & \textbf{UTMOS$\uparrow$} & \textbf{NISQA$\uparrow$} \\
				\midrule
				VoiceCraft & 9.213 & 3.032 & 3.623 & 3.788 & 9.380 & 2.921 & 3.642 & 3.661 \\
				VoiceCraft-X & \textbf{8.289} & \textbf{3.181} & \textbf{3.856} & \textbf{3.901} & \textbf{8.611} & \textbf{3.046} & \textbf{3.892} & \textbf{3.882} \\
				QuarkAudio & 11.275 & 3.121 & 3.562 & 3.722 & 11.973 & 2.941 & 3.626 & 3.645 \\
				\bottomrule
			\end{tabular}
		}
\end{table}

\subsubsection{Audio Event Editing}

\textbf{Evaluation Datasets:}
We select 100 samples from AudioSet and 100 samples from CLAP\_FreeSound 
to simulate the evaluation dataset 
for both the insertion and deletion scenarios.
There is no overlap between these samples and the training dataset.

\textbf{Results:}
It achieves promising event audio editing performance under the text-guided setting, 
attaining an average Multiscale Spectral Distortion (MSD) of 10.08 in the insertion scenario and 
an average MSD of 9.87 in the deletion scenario.

\section{Conclusion}
\label{sec:conclusion}
In this work, we introduce \textbf{\modelraw}, 
a unified framework for audio generation and editing 
based on discrete unified representation. 
By leveraging continuous conditioning and parallel token generation on \textbf{H-Codec}, our approach supports efficient, 
instruction-driven editing of both semantic and acoustic attributes. 
Extensive experiments demonstrate that H-Codec achieves competitive performance across diverse benchmarks, 
the different versions of H-Codec support a broad spectrum of real-world application scenarios. 
We utilize the H-Codec in autoregressive modeling for audio processing and editing, 
where task-specific behavior is controlled through task tokens, enabling diverse operations within a single framework. 
Experimental results verify that our model has ability to tackle various audio processing tasks and 
has potential to perform speech or audio event editing.
We will continue to improve the performance of our current model, 
with a particular focus on speech semantic editing and sound event editing. 
In parallel, we plan to further advance more capable codec variants and 
explore a broader range of audio task types in downstream applications.
\newpage

\bibliography{iclr2026_conference}
\bibliographystyle{iclr2026_conference}

\clearpage 
\appendix
\section*{Appendix} 
\addcontentsline{toc}{section}{Appendix} 

\section{Data Preparation}
\label{appendix:data_preparation}
A data simulation pipeline is designed to synthesis data pairs dynamically during training. 
Considering various types of degradation in the SR task, 
we apply multiple distortions to a speech sample with independent probabilities, 
where the distortion categories and corresponding configurations are shown in ~\Cref{tab:simu}. 
The distortion chain is also applied to the TSE and rTSE modes,
except that the probability of interfering speaker is set to 1.0 and the SIR is uniformly sampled between 
-5 and 5 dB. 
For the LASS mode, we mix the target audio with another randomly selected audio using a SIR ranges from -5 to 20 dB. 
For the VC mode, we leverage a voice conversion model\footnote{https://github.com/myshell-ai/OpenVoice} to perform timbre perturbation using randomly selected target speech and reference speech, 
generating 6k hours of fixed training dataset. 
The perturbed sample is used as input to predict the target speech based on another speech of the target speaker.

For the speech semantic editing (EDIT-S), we employ LLMs to perform text editing and utilize commercial voice 
cloning algorithms to generate corresponding speech. 
Specifically, we first filter samples exceeding 10 words in length from a high-quality ASR training dataset. 
Subsequently, the Qwen3 API is applied to expand and edit the transcribed texts through random operations, 
generating 10 instances of each editing type (insertion, deletion, and substitution). 
Finally, commercial voice cloning interfaces are employed to synthesize speech by using the filtered speech 
samples as reference materials and the edited texts as generation prompts, resulting in approximately 5k hours 
of semantically edited speech corpus.

For the audio event editing (EDIT-A), we adapt the methodology from Recomposer \citep{ellis2025recomposer}. 
Specifically, we first filter audio segments with less than 1 second of silence and containing more than two distinct sound components from the AudioSet dataset as background audio. 
Subsequently, we selected audio samples with captions containing specific sound category keywords from the FreeSound dataset, 
where the keywords are curated from the 521 classes of the AudioSet dataset. 
The classification consistency between the filtered audio and the associated keywords is validated using YAMNet. 
Finally, 20,000 audio samples are extracted from the filtered AudioSet dataset, and each background audio is randomly superimposed with 30 selected 
FreeSound audio samples through event-based audio mixing, resulting in approximately 1k hours of event audio editing training data.

\begin{table}[htbp]
	\centering
	\small
	\caption{Distortion categories and corresponding configurations, 
    where SNR and SIR denote the signal-to-noise ratio and signal-to-interference ratio, respectively.
    }
	\label{tab:simu}
	\begin{tabular}{lcc}
		\toprule
		\textbf{Distortion} & \textbf{Occurrence Probability} & \textbf{Hyperparameters} \\
		\midrule
        \midrule
		Additive Noise & 0.5 & SNR $\sim$ Uniform([-15, 20]) dB \\
		\midrule
		Reverberation & 0.4 & - \\
		\midrule
		Clipping & 0.3 & 
		\begin{tabular}{@{}c@{}}
			Min\_quantile $\sim$ Uniform([0.0, 0.1]) \\
			Max\_quantile $\sim$ Uniform([0.9, 1.0])
		\end{tabular} \\
		\midrule
		Bandwidth Limitation & 0.3 & Cutoff frequencies $\in$ \{2, 4\} kHz \\
		\midrule
		Packet Loss & 0.3 & Loss rate $\sim$ Uniform([0.05, 0.25]) \\
		\midrule
		Interfering Speaker & 0.2 & SIR $\sim$ Uniform([15, 25]) dB \\
		\bottomrule
	\end{tabular}
\end{table}

\section{Evaluation Metrics}
\label{appendix:evaluation_metrics}

\subsection{Codec Metrics}
\label{appendix:codec_metrics}
\textbf{PESQ}~\citep{rix2001perceptual}: 
The perceptual evaluation of speech quality (PESQ) assesses perceptual speech quality by comparing 
the reconstructed speech to the ground-truth target speech signal. 
We employ the wideband PESQ scoring from 1 (poor) to 4.5 (excellent).

\textbf{STOI}~\citep{andersen2017non}: 
The short-time objective intelligibility (STOI) evaluates the intelligibility of speech signals, ranging from 0 to 1. 
The higher STOI score indicates a higher intelligibility and better preservation of the speech content.

\textbf{UTMOS}~\citep{saeki2022utmos}: 
An automatic Mean Opinion Score (MOS) predictor\footnote{https://github.com/tarepan/SpeechMOS} measuring the naturalness of speech. 

\textbf{WER}: 
Word Error Rate (WER) measures the intelligibility of the generated speech by using the automatic speech recognition (ASR) model. 
We utilize a HuBERT-based ASR system\footnote{https://huggingface.co/facebook/hubert-large-ls960-ft} to calculate WER. 

\textbf{SPK-SIM}: 
A WavLM-based speaker verification model\footnote{https://github.com/microsoft/UniSpeech/tree/main/downstreams/speaker\_verification} 
is used to calculate the speaker similarity between the reconstructed speech and target speech.

\textbf{STFT Loss \& Mel Loss}: 
We calculate the L1 loss between the magnitude spectrum of the reconstructed speech and target speech, 
where the STFT is performed using a Hann window with a length of 1024 and a shift of 256.
For the Mel Loss, 100 mel filters are utilized.

\subsection{Audio Task Metrics}
\label{appendix:task_metrics}

\textbf{DNSMOS}~\citep{reddy2022dnsmosp835nonintrusiveperceptual}: 
DNSMOS is a neural network-based MOS estimator\footnote{https://github.com/microsoft/DNS-Challenge/tree/master/DNSMOS} that correlates strongly with human quality ratings. 
It comprises three components: 1) speech quality (\textbf{SIG}), 2) background noise quality (\textbf{BAK}), and 3) overall quality (\textbf{OVRL}). 
Note that for the VC task, DNSMOS scores are calculated by averaging three components. 

\textbf{NISQA}~\citep{mittag2021nisqa}: 
NISQA\footnote{https://github.com/gabrielmittag/NISQA} is a deep learning framework for speech quality prediction. 
We report NISQA for the TSE and VC tasks. 

\textbf{SIM}: 
For the TSE task, we evaluate the speaker similarity using finetuned WavLM-base\footnote{https://huggingface.co/microsoft/wavlm-base-plus-sv} following \citep{LauraTSE}. 
While for the VC task, speaker embeddings are computed using the WavLM TDNN\footnote{https://github.com/microsoft/UniSpeech/tree/main/downstreams/speaker\_verification}. 

\textbf{WER}: 
We utilize the whisper-large-v3\footnote{https://huggingface.co/openai/whisper-large-v3}~\citep{radford2023robust} 
to obtain the transcriptions of converted speech in the VC task, thereby calculating WER with 
the ground-truth text of source speech. 
For the EDIT-S task, whisper-large-v3 
and paraformer-zh\footnote{https://huggingface.co/funasr/paraformer-zh} 
are utilized for English and Chinese test sets, respectively.

\textbf{PLCMOS}~\citep{DienerPSSAC23}: 
A metric\footnote{https://github.com/microsoft/PLC-Challenge/tree/main/PLCMOS} designed to evaluate the quality of speech enhanced by PLC algorithms, 
outputting a single score ranging from 1 to 5 (higher is better). 

\textbf{FAD}~\citep{kilgour2018fr}: 
Fréchet Audio Distance (FAD)\footnote{https://github.com/gudgud96/frechet-audio-distance} measures the quality of generated audio by comparing the statistics of deep features between real and synthesized audio. 
Lower FAD value indicates higher fidelity and better distributional alignment.

\textbf{CLAPScore \& CLAPScore$_A$}~\citep{clap}: 
CLAPScore measures text-audio similarity using joint embeddings from a contrastive language-audio pretraining (CLAP) model\footnote{https://github.com/LittleFlyingSheep/CLAPScore\_for\_LASS}.
While CLAPScore$_A$ evaluates the similarity between the output audio and the target audio. 

\textbf{MSD}~\citep{wang2019neuralsourcefilterwaveformmodels,engel2020ddspdifferentiabledigitalsignal}: 
Multiscale Spectral Distortion (MSD) calculates the
cellwise difference between spectrograms calculated at various 
timere solutions, averaging both linear- and log-domain results.
MSD isessentially a signal-level metric, but is more tolerant
of minor differences in timing when compared to waveform
mean-squared error.

\end{document}